\documentclass[12pt]{elsarticle}

\usepackage{caption}
\usepackage{subcaption}

\usepackage[cmex10]{amsmath,mathtools}
\usepackage{mdwmath}
\usepackage{mdwtab}
\usepackage{amssymb}

\usepackage{algorithmic}
\usepackage{algorithm}
\usepackage{picins}
\usepackage{float}

\newcommand{\algorithmicinput}{\textbf{Input:}}
\newcommand\INPUT{\item[\algorithmicinput]}

\usepackage{epstopdf}
\usepackage{multirow}
\usepackage{color}

\journal{IEEE Transactions on Vehicular Technology}

\begin{document}

\begin{frontmatter}

\title{Wireless Power Transfer and Data Collection in Wireless Sensor Networks}

\author[sutd]{Kai Li}
\ead{kai\_li@sutd.edu.sg}

\author[csiro]{Wei~Ni}

\author[sutd]{Lingjie~Duan}

\author[sydu]{Mehran~Abolhasan}

\author[beihang]{Jianwei~Niu}

\address[sutd]{The Singapore University of Technology and Design, Singapore}

\address[csiro]{Commonwealth Scientific and Industrial Research Organization, Australia}

\address[sydu]{The University of Technology Sydney, Australia}

\address[beihang]{State Key Laboratory of Virtual Reality Technology and Systems, Beihang University, China}

\begin{abstract}
In a rechargeable wireless sensor network, the data packets are generated by sensor nodes at a specific data rate, and transmitted to a base station. Moreover, the base station transfers power to the nodes by using Wireless Power Transfer (WPT) to extend their battery life. However, inadequately scheduling WPT and data collection causes some of the nodes to drain their battery and have their data buffer overflow, while the other nodes waste their harvested energy, which is more than they need to transmit their packets. 
In this paper, we investigate a novel optimal scheduling strategy, called EHMDP, aiming to minimize data packet loss from a network of sensor nodes in terms of the nodes' energy consumption and data queue state information. 
The scheduling problem is first formulated by a centralized MDP model, assuming that the complete states of each node are well known by the base station. This presents the upper bound of the data that can be collected in a rechargeable wireless sensor network. 
Next, we relax the assumption of the availability of full state information so that the data transmission and WPT can be semi-decentralized. The simulation results show that, in terms of network throughput and packet loss rate, the proposed algorithm significantly improves the network performance.
\end{abstract}


\end{frontmatter}

\section{Introduction}
\label{Introduction}
\begin{figure}[htb]
\centering
\includegraphics[width=3.0in]{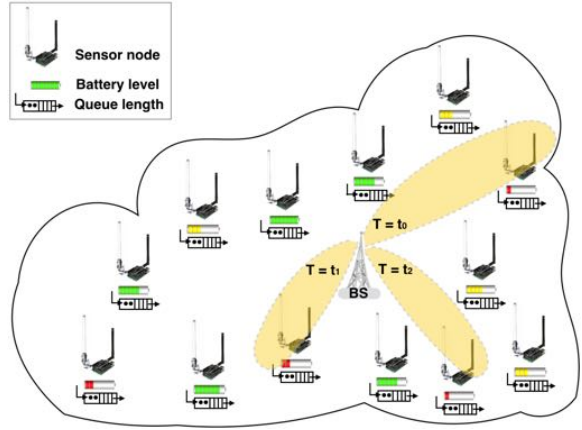}
\caption{\small{Data collection and WPT in wireless powered sensor network.} }
\label{fig_wpt_network}
\end{figure}

Inexpensive sensors capable of significant computation and wireless communications are becoming available. However, sensor nodes are severely constrained by the amount of battery power, limiting the network lifetime and quality of service. Rechargeable Wireless Sensor Network (RWSN) has been extensively studied, where energy is harvested by Wireless Power Transfer (WPT) to recharge battery of a sensor node~\cite{fu2016esync,che2015spatial,xiao2014wireless,sudevalayam2011energy}. 
Figure~\ref{fig_wpt_network} depicts that a number of fixed sensor nodes serve as data sources with sensing ability, for example, monitoring the environment and detecting abnormal events. 
The nodes, equipped with a data communication antenna and a wireless power receiver, generate data packets at an application-specific sampling rate. A base station (BS) is deployed to transfer power to sensor nodes via WPT, and collect sensory data from the nodes~\cite{zhang2016optimal,li2016fair,shu2016near,li2015poster}. Beamforming is used at the BS, either electronically or mechanically. The use of beamforming allows for the concentrated transfer of energy towards the intended node, which overcomes the broadcast nature of radio, and avoids the dispersion and waste of energy~\cite{wang2016adaptively,gong2016optimal,che2015multiantenna}. 
During each time slot (or epoch), one of the nodes is scheduled to transmit data to and harvest energy from the BS. 
To this end, the BS only generates one beam per epoch for both energy transfer and data collection, thereby reducing the overhead of beamforming. 
Especially, the network in Figure~\ref{fig_wpt_network} can be generically extended to a scalable RWSN that is composed of multiple cells, where each cell is covered by one BS. 
The working mode of WPT includes power splitting, and time switching in terms of the WPT receiver structure~\cite{ding2015application}. With time switching, a receiver switches its operation between the power transfer and data transmission over time, while with power splitting, the received signal is split into two streams with one stream used for transferring power and the other stream for transmitting data. 
In particular, we consider a time-switching WPT network in this work. The WPT and data transmission are designed to be in the same radio frequency band, and different time slots so that the nodes only need a single RF chain, which reduces the hardware cost. 
It is critical to schedule the WPT and data collection to extend the battery life of the nodes and to minimize the packet loss. 
Inadequately scheduling WPT and data collection causes some of the nodes to drain their battery and have their data buffer overflow, while the other nodes waste their harvested energy, which is more than they need to transmit their packets. 

In this paper, we aim to optimize the scheduling of WPT and data collection in RWSNs. 
The scheduling problem is first formulated as an off-line finite state Markov Decision Process (MDP), called EHMDP, where transmission order, modulation level, and WPT duration of the nodes are jointly optimized. 
As a result, the packet loss of entire network can be minimized and the battery lifetime of all the nodes are leveraged. 
Note that in EHMDP, the BS needs to have real-time knowledge on the packet arrivals and battery levels of the nodes. This requires the nodes to report prior to every time slot. 
Extending from the centralized scheme, we further propose a semi-decentralized approach that allows the nodes to spontaneously self-nominate for data transmission and energy harvesting. Each sensor node makes its own decision on whether it turns on transmission in a time slot, as well as the modulation level if the transmission is turned on. 
We proposed a transmission probability to impose urgency to the nodes with built-up queues, avoid their transmissions being withheld due to insufficient energy harvested, and in turn, avoid data queue overflows at the nodes and reduce packet losses. Furthermore, an Energy-Queue Aware Transmission (E-QAT) algorithm is proposed in terms of the transition probability of the node. 
In addition, while this paper is structured around WSNs, our framework can be applied to any wireless networks with energy harvesting capabilities. 

The rest of this paper is organized as follows. The related work is summarized in Section~\ref{literature}. Section~\ref{system} presents network structure of RWSN. The centralized MDP model with global information for WPT and data collection is investigated in Section~\ref{mdp_formulate}. In Section~\ref{algorithm}, a semi-decentralized algorithm based on individual energy and data queue is proposed. Simulations and numerical results are presented in Section~\ref{simulation}. Conclusions and future work are given in Section~\ref{cond}.

\section{Related Work}
\label{literature}
Earlier energy-efficient transmission schemes were focused on a single point-to-point link~\cite{nan2016energy,chen2015provisioning,chen2016optimal,cui2017energy,cui2017energy2} or in cellular frameworks~\cite{ni2013new}. In~\cite{nan2016energy}, a string tautening algorithm was proposed to produce the most energy-efficient schedule for delay-limited traffic, first under the assumption of negligible circuit power, and then extended to energy-harvesting powered transmissions with non-negligible transmitter circuit powers~\cite{chen2015provisioning,chen2016optimal}. Later, simple network topologies, such as three-party two-way relay, were considered for the maximization of energy efficiency, first in a half-duplex transmission mode~\cite{cui2017energy} and then in a full-duplex mode~\cite{cui2017energy2}. 
Several studies have considered to integrate WPT technologies into communication networks to achieve simultaneous data transmission and power transfer~\cite{he2013energy,fu2016optimal,zhang2016data,zhao2016optimal,xu2014multiuser,zhou2014wireless,lee2016energy,kang2015full,tekbiyik2013proportional,maravsevic2017max,qin2017joint}. 
In~\cite{he2013energy,fu2016optimal}, a mobile charger is employed to power wireless identification and sensing nodes which support sensing, computing, and communicating. The movement strategy of the mobile charger is formulated, such that the charging latency of all nodes can be minimized while the continuous operation of the sensor nodes can be guaranteed. 
In~\cite{zhang2016data}, the energy harvested by a sensor node is allocated for sensing and data transmitting in a balanced fashion. The sensing rate and routing are optimized, given the dynamic features of network topology. 
In~\cite{zhao2016optimal}, a data sender harvests energy from a power transmitter via WPT in the downlink before transmitting information to a data receiver in the uplink. It is found that data reception performance degrades when the time for energy harvesting and data transmission is unbalanced.   
In~\cite{xu2014multiuser}, power transfer weight and power allocation at the BS are designed to balance the transferred power to the receivers given different SINR requirements. Moreover, the problem of joint information and energy transmission is formulated to give the upper bound of the sum of transferred power. 
Zhou \textit{et al.} present a resource allocation problem for simultaneous data and WPT in downlink multiuser OFDM system, where the users harvest energy and decode data using the same signals received from the BS~\cite{zhou2014wireless}. A tradeoff between the weighted sum-rate of all users and transferred energy is obtained given lower bound of harvested energy on each user and upper bound of total transmission power. 
Moreover, an energy harvesting resource allocation scheme has been investigated with imperfect channel state information in~\cite{lee2016energy}. 
In~\cite{kang2015full}, WPT is considered to support multiple users concurrently, where the BS broadcasts the energy in the network. A frame structure is proposed to allocate time for WPT and data transmission, which achieves the maximum of sum-throughput and minimum of charging and transmission time. 

To achieve a balance between throughput and data collection fairness, a scheduling strategy is proposed to allocate the power transfer among users and the proportion of the time between energy harvests~\cite{tekbiyik2013proportional}. 
In~\cite{maravsevic2017max}, a data rate selection in RWSN is studied, which considers fairness of data rate and WPT duration among the nodes. Moreover, a set of algorithms are developed to obtain a optimal rate assignment with the Water-Filling-Framework under different routing types. 
A WPT system is studied to balance a tradeoff between spectral and energy efficiencies by jointly designing beamforming for WPT and data transmission~\cite{qin2017joint}. Moreover, both WPT and data transmission beamforming are adaptive to channel fading by exploiting the benefits of channel state information. 

However, most of the work in simultaneous WPT and data transmission only focus on improving energy efficiency of WPT. The data lost caused by buffer overflow is not correlatively considered. 

MAC protocol design for scheduling data transmission and power transfer is given in the literature~\cite{kunikawa2015fair,naderi2014rf,nintanavongsa2013medium}. 
It was noted that the power transfer varies depending on surrounding environments. A MAC protocol based on fair polling scheme is studied to achieve a data collection fairness~\cite{kunikawa2015fair}. Moreover, each node contends channel adaptively to their energy harvesting rate. 
In~\cite{naderi2014rf}, a RF-MAC protocol is studied to jointly schedule power transmitters and energy harvesting rates in RWSN. RF-MAC focuses on the amount of energy delivery to the nodes, while eliminating disruption to data communication. 
In~\cite{nintanavongsa2013medium}, multiple WPT transmitters are grouped into different sets based on an estimate of their separation distance from the node to reduce the impact of destructive interference. Moreover, a MAC protocol is presented to transfer power to the nodes on request. 
Unfortunately, the MAC protocols in literature schedule WPT and data communication with the objective of interference elimination, which is different from the problem in this work.

\section{System Design and Structure}
\label{system}
\begin{table}[htb]
    \centering
    \caption{The list of fundamental variables defined in system model}
    \begin{tabular} {|p{1.5cm}|p{5cm}|} \hline
        \bf{Notation} & \bf{Definition} \\ \hline
        	N &  number of sensor nodes  \\ \hline
	i &  node ID  \\ \hline 
	$\mathbf{h}$ &  channel gain between the BS and the node  \\ \hline
	$P^{E}$ &  power that is transferred to the node \\ \hline
	$P_{e}$ &  output power of the BS  \\ \hline
	$L$ &  number of bits of the sensory data packet \\ \hline
	$\epsilon$ &  required BER of the node \\ \hline
	$e_{i}$ &  battery level of node $i$ \\ \hline
	$E$ &  battery capacity of the node  \\ \hline
	$K$ &  the highest battery level of the node  \\ \hline
	$q_{i}$ &  queue length of node $i$ \\ \hline
	$Q$ &  maximum queue length of the node \\ \hline
	$\rho$ &  modulation scheme of the node \\ \hline
	$M$ &  the highest modulation order \\ \hline
	$\lambda$ &  arrival probability of the data packet \\ \hline
	$\mathcal{A}$ &  action set of MDP \\ \hline
	$T_{\lambda}$ & time when a new incoming data packet is put into the queue \\ \hline
	$\widehat{T}$ & time interval of the proposed discrete-time Markov decision process. \\ \hline
   \end{tabular}
\label{tb_variables}
\end{table}
Notations used in this paper are listed in Table~\ref{tb_variables}. 
The network under consideration consists of a BS and $N$ geographically distributed energy-harvesting powered nodes, as illustrated in Fig.~\ref{fig_wpt_network}. 
The BS, connected to persistent power supply, is responsible for remotely charging the nodes using WPT. Equipped with $N_c$ antennas ($N_c\gg 1$), the BS can exploit transmit beamforming techniques to produce a narrow beam to each node. As a result, energy is transferred with improved focus and transfer efficiency. 
The BS is also responsible for collecting sensory data. Additionally, receive beamforming techniques enable the BS to enhance the received signal strength (RSS) and reduce bit error rate (BER). 
In particular, other advanced multi-user beamforming techniques, e.g., zero-forcing beamforming, are not considered in this work since they achieve spatial multiplexing or diversity, but would require real-time feedback on channel state information in most cases. 

Each of the energy-harvesting powered nodes, e.g., node $i$ ($i=1,\cdots,N$), harvests energy from the BS to power its operations, e.g., sensing and communicating. The rechargeable battery of the node is finite with the capacity of $E_i$ Joules, and it overflows if overcharged.
The complex coefficient of the reciprocal wireless channel between the BS and the node is $\mathbf{h}_i$, which is precisely known at both the BS and the node through channel reciprocity. Considering the non-persistent power supply and subsequently limited signal processing capability of the nodes, we let each node be equipped with a single antenna. $\mathbf{h}_i\in \mathcal C^{N_c\times 1}$.
Suppose that the BS employs maximal ratio combining (MRC)~\cite{chen2013energy} to maximize the beamforming gain. The power transferred to node $i$ can be given by
\begin{equation}\label{eq: transfer power}
P_i^E=P_e\|\mathbf{h}_i\|^2,
\end{equation}
where $P_{e}$ is the constant output power of the BS, and $\|\cdot\|$ stands for norm.

Node $i$ also keeps sensing its ambient environment, packetizes and queues (in a first-in-first-out (FIFO) fashion) the sensory data in packets of $L_i$ bits, and transmits the packets to the BS through the wireless channel.
The arrival/queueing process of sensory data at node $i$ is modeled as a random process, where a new packet is put into the FIFO queue at a probability $\lambda_i$ within every $T_{\lambda}$ seconds. 
The node has a finite buffer to accommodate the maximum queue length of $Q_{i}$ packets or $Q_{i}L_{i}$ bits. 
The buffer starts overflowing, once the maximum queue length is reached while the transmission of the sensor is withheld due to insufficient energy harvested. In other words, the arrival rate of sensory data exceeds the departure rate, due to the insufficient energy harvested. To this end, the scheduling algorithm needs to be appropriately designed to assign the sensor sufficient channel and energy resources. 

The modulation scheme that node $i$ uses to transmit packets is denoted by $\rho_i$. $\rho_i\in \{1,2,\cdots,M\}$, where $\rho_i=1,2$, and 3 indicates binary phase-shift keying (BPSK), quadrature-phase shift keying (QPSK), and 8 phase-shift keying (8PSK), respectively, and $\rho_i\geq 4$ corresponds to $2^{\rho_i}$ quadrature amplitude modulation (QAM). $M$ is the highest modulation order.

Suppose that the bit error rate (BER) requirement of node $i$ is $\epsilon_i$, and the MRC is conducted at the BS to maximize the signal-to-noise ratio (SNR). The required transmit power of node $i$ depends on $\rho_i$, $\epsilon_i$, and $\mathbf{h}_i$, as given by~\cite{li2016energy,he2014optimal}
\begin{equation}\label{eq: transmit power}
P_i^D(\rho_i)\approx\frac{\kappa_2^{-1}\ln\frac{\kappa_1}{\epsilon_i}}{\|\mathbf{h}_i\|^2}(2^{\rho_i}-1),
\end{equation}
where $\kappa_1$ and $\kappa_2$ are channel related constants.

For illustration convenience, we consider a homogeneous network, where all the nodes have the same battery size, buffer size, packet length, packet arrival probability, wireless channel, and the BER requirement. The subscript ``$_i$'' is suppressed in $E_i$, $Q_i$, $L_i$, $\lambda_i$, $\mathbf{h}_i$, and $\epsilon_i$. 
However, the proposed scheduling protocols can be extended to a heterogeneous network, where the complexity of the scheduling problem may grow, as the result of an increased number of states. 
Moreover, all the channels involved are assumed to be block-fading, i.e., the channels remain unchanged during each transmission block, and may change from block to block.

\section{Centralized Scheduling and Off-line Optimization}
\label{mdp_formulate}

\subsection{Centralized Scheduling Protocol}
One time slot, also known as a scheduling interval, lasts $\widehat{T}$ mini-slots, during which the node sends a packet to the BS, followed by WPT from the BS to the node. 
In this case, the BS takes the action of choosing a node and deciding the associated modulation order for every time slot. After receiving the packet of the node, the BS starts to transfer energy and it can acknowledge its successful or failed reception of the packet by changing the phase or waveform of the energy signal~\cite{kadrolkar2012variable}. The node discards the packet, if successfully delivered, from its queue, based on the acknowledgement.
Given such centralized coordination, the nodes are activated in a TDMA fashion to transmit sensory data and harvest energy, where one node per time slot. 
We note that there is no collision between the data transmissions of the nodes in the case of centralized scheduling. The receive and transmit beams can also be produced ahead of the actual data transmission or energy transfer.

We note that the BS's actions are not only based on the current network state (i.e, the battery level and queue length of every node $i$, denoted by  $e_i$ and $q_i$, respectively, $i=1,\cdots,N$), but take into account potential influence on future evolutions of the network as well. Particularly, the current action that the BS takes can affect the future battery level and queue length of every node, and in turn, influences the future actions to be taken. Such action taking is a discrete time stochastic control process which is partly random (due to the random and independent arrival/queueing process of sensory data at every node) and partly under the control of the decision-making BS. The action of selecting node and modulation for every time slot can be optimized in a sense that the optimality in regards of a specific metric, e.g., packet loss, can be achieved in long term over the entire stochastic control process (rather than in an individual time slot).

\subsection{Off-line EHMDP Modeling}

As noted earlier, the optimization of the action, i.e., the selection of node and modulation, for every time slot needs to be conducted over the entire stochastic control process of centralized scheduling. The correlation between actions taken in different time slots needs to be captured, and to validate the long-term optimality of the actions.

We consider optimizing the actions to minimize the overall packet loss of the entire WPT powered sensor network. The packet loss can result from buffer overflows at the nodes where data transmissions are withheld due to insufficient energy harvested. The packet loss can also result from unsuccessful data transmissions over wireless fading channels.

It has been known that battery readings are continuous variables with variance difficult to be traced in real-time. The proposed quantization that discretizes the continuous battery readings to $K$ levels can facilitate generating MDP states in EHMDP. Moreover, the accuracy of the quantization can be improved by reducing the quantization interval, but can result in an increased number of MDP states, and hence an increasing complexity of MDP. 
Therefore, to improve the mathematical tractability of the problem and illustration convenience, the continuous battery is discretized into $K$ levels, as $0 <\mathcal {E} <2\mathcal {E}<\cdots<K\mathcal {E}=E$. $e_i\in\{0,\mathcal {E},2\mathcal {E},\cdots,K\mathcal {E}\}$. In other words, the battery level of a node is lower rounded to the closest discrete level. 
The queue length $q_i\in\{0,1,\cdots,Q\}$ is discrete.

The nodes use a small control message to update the BS with their battery level and queue length at the beginning of every time slot. For example, consider a network of 40 nodes, battery level of 100 and queue length of 100 packets, the overhead of one node takes 12 bits, and total overhead is 480 bits, which is much smaller than the size of a data packet. 
Therefore, we assume that both the transmission time and the energy consumption are negligible. We also assume that the BS receives all the necessary information timely and correctly. In other words, the information required to take optimal actions is assumed to be globally available in the case of centralized scheduling. 
This assumption is reasonable, because the control messages typically adopt reliable, low-order modulation schemes, e.g., BPSK (sometimes with repetition coding). 

Such centralized scheduling can be formulated as a discrete-time Markov decision process (MDP) with time interval of $\widehat{T}$, where each state, denoted by $\mathcal{S}_\alpha$, represents the battery levels and queue lengths of all the nodes in the network, i.e., $\{(e_{\alpha,n},q_{\alpha,n}), n=1,\cdots,N\}$. The size of the state space (i.e., the number of such states) is $\big(K(Q+1)\big)^N$.
The action to be taken during a transition, denoted by $\mathcal A$, is to select the node to be activated and specify its modulation. $\mathcal{A}\in\Big\{(i,\rho_i):i=1,\cdots,N,\rho_i\in \{1,\cdots,M\}\Big\}$, where $M$ denotes the highest modulation level. 
The size of the action set is $NM$.

Furthermore, $\mathcal{A}$ can be reduced to only consist of the selected node, i.e., the $n$-th action $\mathcal{A}_n=\{n\}$. The maximum energy that can be harvested into the battery of the selected node $n$ can be given by
\begin{equation}
\begin{aligned}
\Delta \mathcal{E}_n
=&\bigg\lfloor
(\widehat{T}-\frac{L}{\rho^{\star}_{n} W}) \frac{P_{e} \|\mathbf{h}_{n}\|^2}{\mathcal{E}}-\\
&\qquad \qquad \frac{L\kappa_2^{-1}\ln(\frac{\kappa_1}{\epsilon})}{
\|\mathbf{h}_{n}\|^2\rho^{\star}_{n} W\mathcal{E}}
\big(2^{\rho^{\star}_{n}}-1\big)\bigg\rfloor  \mathcal{E}
\label{eq_deltaE}
\end{aligned}
\end{equation}
where $\rho^{\star}_{n}$ denotes the optimal modulation scheme of node $n$, which is obtained by Equation~(\ref{eq_rho_final}) (See Appendix). 
The actions of the MDP can be optimized by judiciously deriving the transition probabilities and formulating a dynamic programming problem, and solved by using dynamic programming techniques~\cite{puterman2014markov}. Details will be provided shortly in this section.

Note that the centralized scheduling is ideal and can only be conducted off-line, under the assumptions of negligible signaling time and energy, as well as the resultant global availability of the required information. The optimal actions of the MDP are inapplicable to practical environments with non-negligible signaling time and energy consumption. Nevertheless, the MDP (or its optimal actions) provides the lower bound of packet loss to any scheduling designs of the WPT powered sensor networks. Moreover, the MDP also lays foundations to and eases the understanding of the on-line optimization of semi-decentralized scheduling, as will be described in detail in the next section.

%
%
%
%
%

\subsection{Transition Probability and Packet Loss} 
Given the action $\mathcal A_k=\{k\}$ ($1\leq k \leq N$), the transition probability of the MDP, from state $\mathcal{S}_{\alpha}$ to $\mathcal{S}_{\beta}$ ($1\leq (\alpha,\beta) \leq K^N(Q+1)^N$), can be given by
\begin{equation}\label{eq: transit prob}
\begin{aligned}
\Pr\Big\{\mathcal{S}_{\beta}\Big|\mathcal{S}_{\alpha},\mathcal{A}_k\Big\}
=\Pr\Big\{(e_{\beta,k},q_{\beta,k})\Big|(e_{\alpha,k},q_{\alpha,k}),k\in \mathcal{A}_k\Big\}& \\
\times\prod_{n=1, n\notin\mathcal{A}_k}^N\Pr\Big\{(e_{\beta,n},q_{\beta,n})\Big|(e_{\alpha,n},q_{\alpha,n}),n\notin \mathcal{A}_k\Big\}&, 
\end{aligned}
\end{equation}
where the two parts of the RHS are specified in \eqref{eq: transit prob 1&2}. 

Specifically, \eqref{eq: transit prob 1} corresponds to the selected node $k$, where the first case is that the queue of the node increases due to the failed transmission of a packet and the arrival of a new sensory packet. The second case is that the queue decreases due to a successful transmission and no new packet arrival. The third case is that the queue does not change, which is due to either a successful transmission and a new packet arrival, or a failed transmission and no new packet arrival. The battery level of the selected node all increases by $\Delta \mathcal{E}_k$ in the three cases, given the optimized modulation order $\rho_k$.

\eqref{eq: transit prob 2} corresponds to the unselected nodes $n\neq k$, where the first case is that the queue of the node increases due to a new sensory packet arrival. The second case is that the queue does not change, without a new packet arrival. The battery level of the node does not change, since the node is not selected to harvest energy.

\begin{figure*}
\begin{subequations}\label{eq: transit prob 1&2}
\begin{align}
\Pr\Big\{(e_{\beta,k},q_{\beta,k})\Big|(e_{\alpha,k},q_{\alpha,k}),k\in \mathcal{A}_k\Big\}=&\left\{
\begin{array}{l}
\Big(1-(1-\epsilon)^L\Big)\lambda , \\ \qquad~~\text{if }e_{\beta,k}=e_{\alpha,k}+\Delta \mathcal{E}_k \text{ and } q_{\beta,n} = q_{\alpha,n}+1;\\
(1-\epsilon)^L(1-\lambda) , \\  \qquad~~\text{if }e_{\beta,k}=e_{\alpha,k}+\Delta \mathcal{E}_k \text{ and } q_{\beta,n} = q_{\alpha,n}-1;\\
\Big(1-(1-\epsilon)^L\Big)(1-\lambda)+(1-\epsilon)^L\lambda, \\
\qquad~~\text{if }e_{\beta,k}=e_{\alpha,k}+\Delta \mathcal{E}_k \text{ and } q_{\beta,n} = q_{\alpha,n};\\
0 ,  \qquad\qquad\qquad\qquad\qquad\text{otherwise.}\\
\end{array}
\right.\label{eq: transit prob 1}\\
\Pr\Big\{(e_{\beta,n},q_{\beta,n})\Big|(e_{\alpha,n},q_{\alpha,n}),n\notin \mathcal{A}_k\Big\}=&\left\{
\begin{array}{ll}
\lambda , & \text{if }e_{\beta,n}=e_{\alpha,n} \text{ and } q_{\beta,n} = q_{\alpha,n}+1;\\
1-\lambda , & \text{if }e_{\beta,n}=e_{\alpha,n} \text{ and } q_{\beta,n} = q_{\alpha,n};\\
0 , & \text{otherwise.}\\
\end{array}
\right.\label{eq: transit prob 2}
\end{align}
\end{subequations}
\hrulefill
\end{figure*}

The packet loss, resulting from buffer overflow during the transition, can be given by
\begin{equation}\label{eq: transit reward}
\begin{aligned}
R\Big\{\mathcal{S}_\beta\Big|\mathcal{S}_\alpha,\mathcal{A}_k\Big\}
=R\Big\{(e_{\beta,k},q_{\beta,k})\Big|(e_{\alpha,k},q_{\alpha,k}),k\in \mathcal{A}_k\Big\}&\\
+ \sum_{n=1, n\notin\mathcal{A}_k}^NR\Big\{(e_{\beta,n},q_{\beta,n})\Big|(e_{\alpha,n},q_{\alpha,n}),n\notin \mathcal{A}_k\Big\}&;
\end{aligned}
\end{equation}
\begin{subequations}
\begin{align}
R\Big\{(&e_{\beta,k},q_{\beta,k})\Big|(e_{\alpha,k},q_{\alpha,k}),k\in \mathcal{A}_k\Big\}=
\left\{
\begin{array}{ll}
\Big(1-(1-\epsilon)^L\Big)\lambda, & \text{if }q_{\alpha,k}=q_{\beta,k}=Q;\\
0, &\text{otherwise.}
\end{array}
\right. \label{eq: reward 1}\\
R\Big\{(&e_{\beta,n},q_{\beta,n})\Big|(e_{\alpha,n},q_{\alpha,n}),n\notin \mathcal{A}_k\Big\}=
\left\{
\begin{array}{ll}
\lambda, & \text{if }q_{\alpha,n}=q_{\beta,n}=Q;\\
0, &\text{otherwise.}
\end{array}
\right.\label{eq: reward 2}
\end{align}
\label{eq_rewards_2}
\end{subequations}

The first case of \eqref{eq: reward 1} is the probability that the buffer of the selected node $k$ overflows. Specifically, the new generated packet at State $\mathcal{S}_{\beta}$ with the arrival probability $\lambda$ has to be dropped since the node $k$ failed to transmit the packet at State $\mathcal{S}_{\alpha}$ given $Q$ packets in the queue. Similarly, the first case of \eqref{eq: reward 2} is the probability that the buffer of an unselected node $n\neq k$ overflows.

\subsection{MDP Solver}
\label{optimization}
Generally, value iterations, policy iterations, and linear programming are three major methods for computing optimal actions for MDP with expected total discounted rewards. The actions of the MDP are optimized by using dynamic programming techniques~\cite{puterman2014markov}. 
Particularly, in this paper, the value iteration algorithm~\cite{arruda2010toward,sun2008constrained} is applied. The value iteration algorithm computes the optimal cost function by assuming first a one-stage finite horizon, then a two-stage finite horizon, and so on. The cost functions are computed to converge in the limit to the optimal cost function. Therefore, the policy associated with the successive cost functions converges to the optimal policy in a finite number of iterations. 
In addition, the reason is that policy iteration requires policy evaluation at each of iterations, which is a protracted iterative computation involving multiple sweeps through the state set. Linear programming requires the MDP problem can be solved in a number of arithmetic operations polynomial in N and M, which run extremely slowly in practice. However, note that policy iteration and linear programming are also adoptable to obtain the optimal actions of the MDP problem. 

We define that a policy $\pi$ is a mapping from states to actions, and the set of all the policies is given as $\Pi$. Moreover, if the policy is independent of the current stage, it is said to be stationary. The goal of the algorithm is to minimize the expected total packet loss which is denoted as $v(S_{\beta})$, 
\begin{equation}
v(S_{\beta}) = \min_{\pi \in \Pi} \left\{E^{\pi}_{S}\sum^{\infty}_{t=1} \omega^{t-1} R\Big\{\mathcal{S}_\beta\Big|\mathcal{S}_\alpha,\mathcal{A}_k\Big\}\right\}
\label{eq_max_1}
\end{equation}
where $\omega (\omega \in [0, 1])$ denotes the discount factor for future states.  
To find the action $k$ to calculate $v(S_{\beta})$, we evaluate for each action $k \in \mathcal{A}_k$ and select the actions that achieve Equation~\eqref{eq_max_1}. Thus, we have
\begin{equation}
\begin{array}{c l}
v(S_{\alpha}) =& \min_{k \in \mathcal{A}_k} \{R\Big\{\mathcal{S}_\beta\Big|\mathcal{S}_\alpha,\mathcal{A}_k\Big\} + \sum_{\mathcal{S}_\beta \in {\cal S}}\omega Pr\Big\{\mathcal{S}_\beta\Big|\mathcal{S}_\alpha,\mathcal{A}_k\Big\} v(\mathcal{S}_\beta) \}
\end{array}
\label{eq_max_2}
\end{equation}

Therefore, the optimal action $k$ which satisfies Equation~\eqref{eq_max_2} can be given by 
\begin{equation}
k = \arg \min_{k \in \mathcal{A}_k} \{R\Big\{\mathcal{S}_\beta\Big|\mathcal{S}_\alpha,\mathcal{A}_k\Big\} + \sum_{\mathcal{S}_\beta \in {\cal S}} \omega Pr\Big\{\mathcal{S}_\beta\Big|\mathcal{S}_\alpha,\mathcal{A}_k\Big\} v(S_{\beta}) \} 
\end{equation} 
Note that $k$ may not be unique, but at least one minimizing action exists due to a finite $\mathcal{A}_k$. Specifically, if $k$ has an unique value, it is the optimal decision in state $S_{\alpha}$. 
If not, choosing any minimizing action can achieve the minimum expected packet loss. 

The Value Iteration Algorithm is summarized in Algorithm~\ref{alg_via}.
\begin{algorithm}[t]
\caption{Value Iteration Algorithm}
\label{alg_via}
\begin{algorithmic}[1]
\STATE{\textbf{1. Initialize}: }
\STATE{$v^{0}(S_{\alpha}) = 0$ for the state $S_{\alpha}$, $t \in [0, \infty]$, specify $\varepsilon > 0$, and set w = 0}
\STATE{\textbf{2. Iteration}: }
\WHILE{$S_{\alpha} \in {\cal S}$}
\STATE{$v^{w+1}(S_{\alpha}) = \min_{k \in \mathcal{A}_k} \{R\Big\{\mathcal{S}_\beta\Big|\mathcal{S}_\alpha,\mathcal{A}_k\Big\} + \sum_{\mathcal{S}_\beta \in {\cal S}}\omega Pr\Big\{\mathcal{S}_\beta\Big|\mathcal{S}_\alpha,\mathcal{A}_k\Big\} v(\mathcal{S}_\beta) \}$}
\IF{$||v^{w+1} - v^{w}|| < \varepsilon (1 - \omega)/(2\omega)$}
\FOR{$S_{\alpha} \in {\cal S}$}
\STATE{$k = \arg \min_{k \in \mathcal{A}_k} \{R\Big\{\mathcal{S}_\beta\Big|\mathcal{S}_\alpha,\mathcal{A}_k\Big\} + \sum_{\mathcal{S}_\beta \in {\cal S}} \omega Pr\Big\{\mathcal{S}_\beta\Big|\mathcal{S}_\alpha,\mathcal{A}_k\Big\} v(S_{\beta}) \}$}
\ENDFOR
\ENDIF
\ENDWHILE
\end{algorithmic}
\end{algorithm}
Since the running time for each iteration is $O(MN^{2})$, the presented value iteration algorithm is polynomial given the polynomial total number of required iterations.

Note that the centralized scheduling is ideal and can only be conducted off-line, under the assumptions of negligible signaling time and energy, as well as the resultant global availability of the required information. The optimal actions of the MDP are inapplicable to practical environments with non-negligible signaling time and energy consumption. Nevertheless, the MDP (or its optimal actions) can provide the lower bound of packet loss to any scheduling designs of the WPT powered sensor networks. The MDP also lays foundations to and eases the understanding of the on-line optimization of semi-decentralized scheduling, as will be described in detail in the next section.

\section{Semi-Decentralized Energy-Queue Aware Data Transmission}
\label{algorithm}
Note that in EHMDP, the BS needs to have real-time knowledge on the packet arrivals and battery levels of the nodes. This requires the nodes to report prior to every time slot. 
Extending from the centralized scheme, we further propose a semi-decentralized approach that allows the nodes to spontaneously self-nominate for data transmission and energy harvesting. Each sensor node makes its own decision on whether it turns on transmission in a time slot, as well as the modulation level if the transmission is turned on. 
The decisions/actions that node $i$ takes are based on the current battery level and queue length of its own, i.e., $e_{i}$ and $q_{i}$, as other nodes' battery and queue statues (i.e., $e_j$ and $q_j$, for any $j\neq i$) are unavailable to node $i$ in distributed networks. 
Collisions occur under such distributed scheduling, due to the independent actions that each node takes.  

\subsection{Protocol Design}
To alleviate the collisions, for every node $i=1,\cdots,N$, we propose to design a transmission probability, denoted by $p_i\in [0,1]$, at which the node is activated to transmit a packet and then harvest energy in the current time slot. Apparently, $p_i$ needs to depend on $e_i$ and $q_i$, which is given as $f(e_{i}, q_{i})$. Our design philosophy is to design $p_i$ to monotonically increase with $q_i$ and decrease with $e_i$. The purpose of such design is to impose urgency to the nodes with built-up queues, avoid their transmissions being withheld due to insufficient energy harvested, and in turn, avoid buffer overflows at the nodes and reduce packet losses. 
Note that the definition of $p_i$ is generic, which can be given by any classical probability distributions. Especially, Exponential, Sigmoid, and Gamma distributions are considered as three design examples of $p_i = f(e_{i}, q_{i})$ in this paper. The reason is that they present variant $p_i$ regarding to the $e_i$ and $q_i$ so that the impact of $p_i$ on the proposed protocol can be observed. 
Specifically, the three examples of $p_i$ are given by 
\begin{equation}
\begin{aligned}
\text{Exponential design}: p_i &= (1-e^{-\kappa_qq_i})e^{-\kappa_ee_i}\\
\text{Sigmoid design}: p_i &= \sin(\frac{\pi}{2}\frac{q_i}{Q})\cos(\frac{\pi}{2}\frac{e_i}{E})\\
\text{Gamma design}: p_i &= \frac{1}{\Gamma(\varrho)}\gamma(\varrho,\frac{1}{\theta}\frac{q_{i}}{e_{i}})\\
\end{aligned}
\label{eq_probTx}
\end{equation}
where for exponential distribution, $\lambda$ denotes the rate parameter; for Gamma distribution, $k > 0$ is the shape value, and $\theta > 0$ is the scale value. $\Gamma(\varrho)$ is the Gamma function, and $\gamma(\varrho, \frac{1}{\theta}\frac{q_{i}}{e_{i}})$ is the lower incomplete gamma function. 

Other techniques to alleviate collisions in distributed scheduling include channel-sensing multiple access/collision-avoidance (CSMA/CA) and exponential back-off retransmission which are widely used in IEEE 802.11/802.15 distributed networks~\cite{chaturvedi2016design,li2014kappa}. 
An extended CSMA-based protocol, namely, Q-CSMA, is able to schedule nodes based on their queue lengths~\cite{ni2012q}. However, these techniques are not suitable for energy-restrained sensor nodes, because CSMA/CA requires non-negligible energy to continuously sense the availability of the channel and would quickly drain the batteries. 

Under distributed scheduling, the BS becomes the passive data collector. Consider an ideal case, in a time slot where only a single node transmits a packet (collision-free), the BS starts by detecting the PHY/MAC header (i.e., the pilot signal) of the packet, then generates a receive beam to the node, and continues to receive the sensory data part of the packet. Once the node completes transmitting the packet, the BS forms a transmit beam to transfer energy to the sensor node over the remaining part of the time slot.
The BS can acknowledge its successful or failed reception of the packet by changing the phase or waveform of the energy signal. The node discards the packet, if successfully delivered, from its queue, based on the acknowledgement. 

Consider multiple nodes have their transmissions collided in a time slot, the BS stays idle, since it can by no means identify the activated nodes. 
We denote ${\Pr}^{c}_{k}$ as the packet collision probability of the selected node $k$ by our MDP model, and it is given by~\cite{shrestha2011markov}
\begin{equation}
{\Pr}^{c}_{k} = 1 - \prod_{n\neq k}(1-p_n )
\label{eq_probCollision}
\end{equation}

Equation~(\ref{eq_transprob_collision_node}) depicts the probability of node $k$ that transmits data with packet collision probability and harvested energy. 
In the first case, the battery level of selected node $k$ increases by $\Delta \mathcal{E}_{k}$, and the queue length also increases, which indicates that a new packet arrives and the packet of State $\alpha$ is transmitted successfully with no collision. 
The second case is that the queue length of selected node $k$ decreases since there is no new packet arrival and the packet of State $\alpha$ is transmitted successfully, and the battery level increases via energy harvesting of WPT. 
The third cases is that the queue length of the node $k$ does not change. The battery level increases at State $\beta$, which denotes that a new packet arrives while the sensory packet of State $\alpha$ is transmitted successfully. 
The fourth case is that the queue length of selected node $k$ remains the same. However, the battery level decreases by $\Delta \nu_k$. Therefore, in this case, there is no new packet arrival, moreover, the transmission at State $\alpha$ has packet collision, which requests a retransmission. 
In the last case, the queue length of selected node $k$ decreases, and the battery level also drops by $\Delta \nu_k$, which indicates that no new packet arrival and the transmitted packet at State $\alpha$ collides. 

Note that a key difference between the probability of the selected node $k$ in Equation~\eqref{eq_transprob_collision_node} and the one in~\eqref{eq: transit prob 1} and~\eqref{eq: transit prob 2} is due to the potential packet collision in the semi-decentralized approach. As a result, the probability formulated in~\eqref{eq_transprob_collision_node} comprehensively considers transmit probability $p_n (n \neq k)$ and the packet collision probability ${\Pr}^{c}_{k}$. 

Given a slot-based data transmission and energy harvesting, the BS needs to broadcast a beacon periodically to all the nodes for network synchronization. $p_n$ ($n \in [1, N]$ and $n \neq k$) can also be broadcast along with the beacon. As a result, the collision probability can be evaluated at each individual node $k$ by using~(\ref{eq_probCollision}). 

\begin{figure*}[!t]
\begin{equation}
\begin{aligned}
\Pr\Big\{(e_{\beta,k},q_{\beta,k})&\Big|(e_{\alpha,k},q_{\alpha,k}),k\in \mathcal{N}\Big\}_{col}=\\
&\left\{
\begin{array}{l}
\big(1-(1-\epsilon)^L\big)\lambda\prod_{n\neq k}(1-p_n),  \\ \qquad~~\text{if }e_{\beta,k}=e_{\alpha,k}+\Delta \mathcal{E}_k \text{ and } q_{\beta,n} = q_{\alpha,n}+1;\\
(1-\epsilon)^L(1-\lambda)\prod_{n\neq k}(1-p_n ), \\ \qquad~~\text{if }e_{\beta,k}=e_{\alpha,k}+\Delta \mathcal{E}_k \text{ and } q_{\beta,n} = q_{\alpha,n}-1;\\
\Big(\big(1-(1-\epsilon)^L\big)(1-\lambda)+(1-\epsilon)^L\lambda\Big)\prod_{n\neq k}(1-p_n ), \\
\qquad~~\text{if }e_{\beta,k}=e_{\alpha,k}+\Delta \mathcal{E}_k \text{ and } q_{\beta,n} = q_{\alpha,n};\\
\Big(\big(1-(1-\epsilon)^L\big)(1-\lambda)+(1-\epsilon)^L\lambda\Big){\Pr}^{c}_{k}, \\ \qquad~~\text{if }e_{\beta,k}=e_{\alpha,k}-\Delta \nu_k \text{ and } q_{\beta,n} = q_{\alpha,n};\\
(1-\epsilon)^L(1-\lambda){\Pr}^{c}_{k}, \\ \qquad~~\text{if }e_{\beta,k}=e_{\alpha,k}-\Delta \nu_k \text{ and } q_{\beta,n} = q_{\alpha,n}-1;\\
0 ,  \qquad~~\text{otherwise.}\\
\end{array}
\right.
\end{aligned}
\label{eq_transprob_collision_node}
\end{equation}
\hrulefill
\end{figure*}

\subsection{Energy-Queue Aware Data Transmission}
We propose an Energy-Queue Aware Transmission (E-QAT) algorithm. Specifically, node $k$ first calculates the transition probability that is given by~\eqref{eq_transprob_collision_node}. In particular, the transition probability is formulated based on the node's packet collision probability, variation of its battery level, harvested energy, and data queue state information. 
If $\Pr\Big\{(e_{\beta,k},q_{\beta,k})\Big|(e_{\alpha,k},q_{\alpha,k}),k\in \mathcal{N}\Big\}_{col}$ is lower than a given threshold, $\Pr^{0}_{thresh}$, the node does not transmit the packet. 
In particular, $\Pr^{0}_{thresh}$ can be determined using collision-based medium access techniques, e.g., the one presented in~\cite{tay2004collision}. 
Then, the transmission probability of node $k$ is updated by $p^{f}_{k} = \min\{1, (1 + \alpha)^{f}p_{k}\}$, where $f$ is the number of frames during which the node failed in data transmission and $\alpha$ is the indicator parameter that can be given by~\cite{chaturvedi2016design}. 
Node $k$ transmits its data packet only when $\Pr\Big\{(e_{\beta,k},q_{\beta,k})\Big|(e_{\alpha,k},q_{\alpha,k}),k\in \mathcal{N}\Big\}_{col} \geq \Pr^{0}_{thresh}$. If the packet is successfully received by the BS, the algorithm continues to process the next data transmission, and $p_{k}$ returns to $f(e_{k}, q_{k})$. 
However, if the data packet collision happens, node $k$ has to back off a random time to retransmit, meanwhile, the transmission probability is correspondingly increased by $p^{f}_{k} = \min\{1, (1 + \alpha)^{f}p_{k}\}$. 

The detail is shown in Algorithm~\ref{alg_eqat}. Moreover, note that the modulation scheme $\rho_{k}$ has been decided by Equation~(\ref{eq_rho}) in Section~\ref{mdp_formulate}. Since the sensor nodes have independent $e_{k}$ and $q_{k}$, the $p^{f}_{k}$ is independently updated in E-QAT, which significantly reduces the collision probability. 

As discussed, the off-line EHMDP protocol and the E-QAT algorithm provide solutions to scheduling data transmission and WPT in centralized and semi-decentralized networks, respectively. In terms of performance, EHMDP provides a lower bound for the overall packet loss and a upper bound for throughput, at the cost of real-time signaling and centralized coordination. Moreover, the transition probability formulation in EHMDP, i.e.,~\eqref{eq: transit prob 1} and~\eqref{eq: transit prob 2}, jointly considers the battery level and queue length of the node. This inspires the design of transmit probability with collision in the E-QAT algorithm, i.e.,~\eqref{eq_transprob_collision_node}, where the nodes can self-nominate for data transmission and energy harvesting based on the current battery level and queue length of its own. 

\begin{algorithm}[htb]
\caption{E-QAT algorithm}
\label{alg_eqat}
\begin{algorithmic}[1]
\INPUT $\Pr\Big\{(e_{\beta,k},q_{\beta,k})\Big|(e_{\alpha,k},q_{\alpha,k}),k\in \mathcal{N}\Big\}_{col}$, \; $p_{k}$
\STATE{\textbf{Repeat:}}
\IF{$\Pr\Big\{(e_{\beta,k},q_{\beta,k})\Big|(e_{\alpha,k},q_{\alpha,k}),k\in \mathcal{N}\Big\}_{col} < \Pr^{0}_{thresh}$}
\STATE{node $k$ does not transmit.}
\STATE{$p^{f}_{k} \gets \min\{1, (1 + \alpha)^{f}p_{k}\}$.}
\ELSE
\STATE{node $k$ transmits the data packet.}
\IF{the packet is collided}
\STATE{$p^{f}_{k} \gets \min\{1, (1 + \alpha)^{f}p_{k}\}$.}
\STATE{back off a random time to retransmit.}
\ELSE
\STATE{$p_{k} \gets f(e_{k}, q_{k})$}
\ENDIF
\ENDIF
\end{algorithmic}
\end{algorithm}

\section{Numerical Evaluation}
\label{simulation}
The simulations prototyping our proposed scheduling are investigated in this section. We compare the performance between our proposed EHMDP and E-QAT, and existing scheduling strategies. We present extensive results showing the performance of E-QAT with different packet collision probability of the scheduled node $k$. Moreover, we explain how $Pr^{c}_{k}$ varies with the MDP states given different probability distribution of $p_{i} (i \in [1,N])$. 

\subsection{Simulation Configuration}
The sensor nodes ($N \in [10, 40]$) are deployed in the range of 50 meters, which are all one-hop away from the BS. 
The node has the maximum discretized battery level $E = 5$ and queue length $Q = 6$, and the highest modulation $M$ is 5. Each data packet has a payload of 32 bytes, i.e., $L = 256$. Moreover, the data transmission period is given as a sequence of time slots where each node generates one data packet and put into the queue. 
For the $P_i^D(\rho_i)$ in Equation~(\ref{eq: transmit power}), the two constants, $\kappa_{1}$ and $\kappa_{2}$ are 0.2 and 3, respectively. We set the target $\epsilon = 0.05\%$ for the numerical results, i.e., the transmitted bits have an error no more than 0.05, however, this value can be configured depending on the traffic type and quality-of-service (QoS) requirement of the sensory data. 
Furthermore, the transmit power of WPT transmitter at BS is 3 watts, and power transfer efficiency, $\delta$, is set to 0.4. 

We simulate the proposed EHMDP in a centralized network, and E-QAT in a distributed network. In either of the scenarios, a random scheduling strategy and a greedy algorithm are simulated for performance comparison. 
In centralized RWSN, the first algorithm, referred to as ``Full Queue (FQ)'', is a greedy algorithm, where the scheduling is based on the data queue length of node. The node with full queue has a higher priority to transmit data and harvest power. 
The second algorithm is named as ``Random Selection (RS)'', where the BS randomly selects one node to transmit data and transfer power. 
In the network with distributed scheduling, the node transmits data and harvests energy based on the local MDP states. The first strategy, referred to as ``Decentralized FQ (D-FQ)'', is a distributed algorithm based on the data queue, where the node with a full queue transmits. The second strategy is named as ``Random Contention (RC)'', where the node transmits data in a random-contention fashion. Note that the BS only transfers power to the node that wins the contention and completes data transmission in the slot. However, due to the distributed scheduling, the packets collide when multiple nodes are in the same MDP state. 

\subsection{Performance Evaluation}
\subsubsection{Centralized and distributed scheduling}
\begin{figure}[htb]
\centering
\includegraphics[width=4.5in]{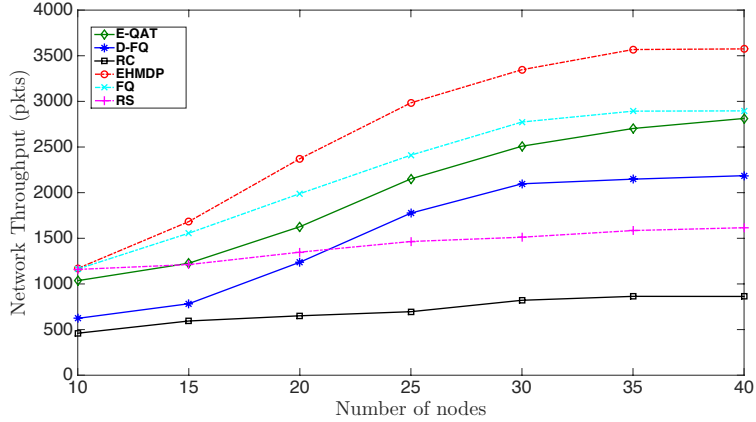}
\caption{\small{A comparison of network throughput using different scheduling strategies in the centralized and distributed RWSN.}}
\label{fig_throughput_ehmdp}
\end{figure}

The network throughput of aforementioned six scheduling strategies are presented in Figure~\ref{fig_throughput_ehmdp}. For centralized scheduling algorithms, EHMDP, FQ, and FS, they have similar throughput when there are 10 nodes in the network. However, from $N=15$ to $N=40$, EHMDP outperforms the other three existing centralized scheduling algorithms. EHMDP achieves 17\%, and 52\% higher throughput than FQ and RS when $N=40$. For semi-decentralized scheduling, $p_{i} (i \in [1,N])$ is given by Sigmoid distribution in Equation~(\ref{eq_probTx}). It is also observed that the throughput of E-QAT is higher than D-FQ and RC by 21\% and 68\%, respectively. 

Furthermore, the throughput of EHMDP is more than E-QAT by around 20\%. The reason is that the state information of each node is known by the BS, which enables the centralized scheduling of the node and modulation for every time slot. 

\begin{figure}[htb]
\centering
\includegraphics[width=4.5in]{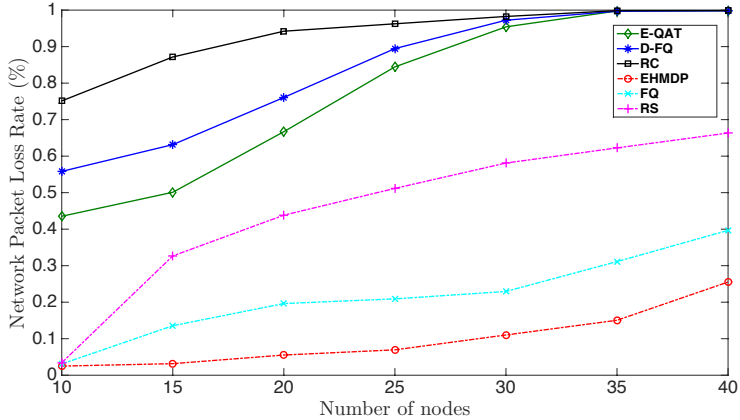}
\caption{\small{A comparison of network packet loss rate by our EHMDP and E-QAT, and the typical scheduling strategies.}}
\label{fig_pktloss_ehmdp}
\end{figure}

In general, the throughput of distributed scheduling algorithms is lower than the centralized ones, which is caused by the packets collision. This issue is also observed by Figure~\ref{fig_pktloss_ehmdp}, which presents the network packet loss rate. The packet loss rate of the centralized algorithms is generally lower than the distributed ones. Specifically, EHMDP achieves 42\%, 57\%, and 72\% less packet loss than E-QAT, D-FQ, and RC when $N=10$. Moreover, the packet loss rate of E-QAT is less than D-FQ and RC output by 16\% and 33\%. We also see that the packet loss rate of the distributed algorithms grows to 100\% with an increase of number of nodes since more time slot contentions and higher $\text{Pr}^{c}_{k}$. 
Moreover, the network throughput and packet loss achieved by E-QAT are much closer to those of the optimal EHMDP, as compared to the other distributed protocols. 
Note that in EHMDP, the BS needs to have real-time knowledge on the packet arrivals and battery levels of the nodes. This requires the nodes to report prior to every time slot. 
However, the semi-decentralized approach allows the nodes to spontaneously self-nominate for data transmission and energy harvesting. Each sensor node makes its own decision on whether it turns on transmission in a time slot, as well as the modulation level if the transmission is turned on. 

\subsubsection{Impact of $p_{i}$ to E-QAT}
\begin{figure}[htb]
\centering
\includegraphics[width=4.5in]{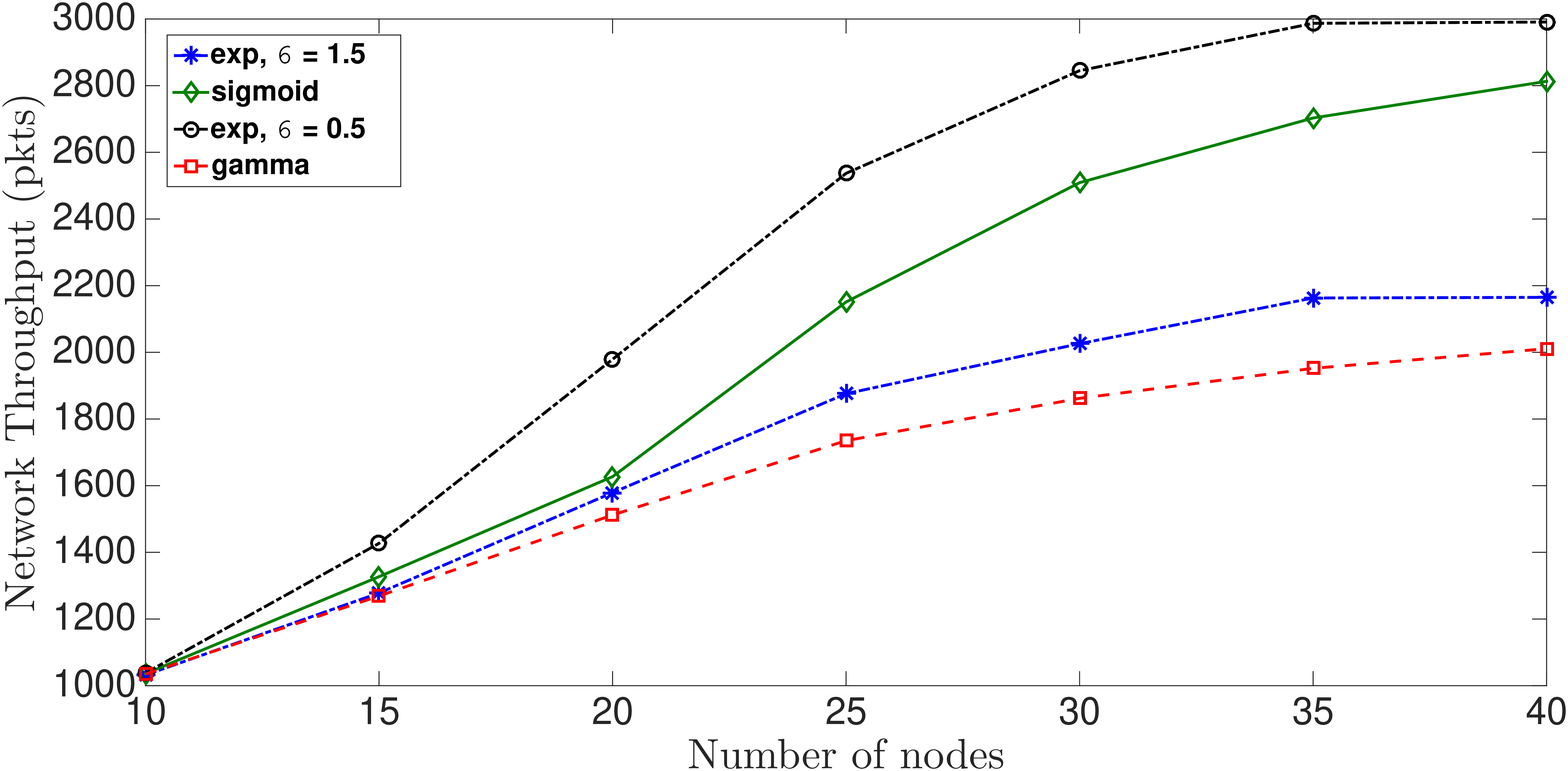}
\caption{\small{Network throughput with different the packet transmission probability in terms of $p_{i} (i \in [1,N])$.}}
\label{fig_dehmdp_throughput_distribution}
\end{figure}

\begin{figure}[htb]
\centering
\includegraphics[width=4.5in]{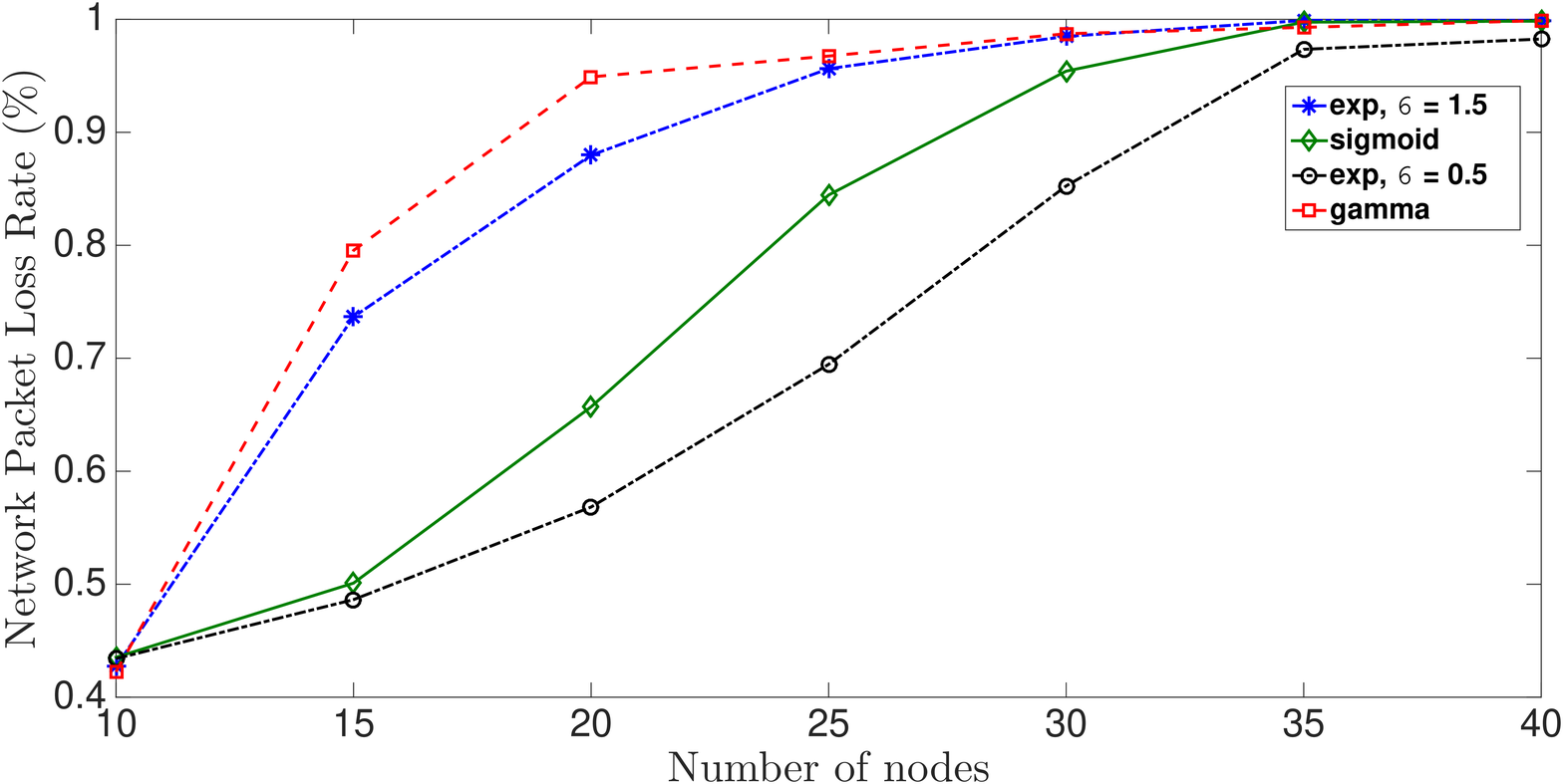}
\caption{\small{Network packet loss with different $p_{i} (i \in [1,N])$.}}
\label{fig_dehmdp_pktloss_distribution}
\end{figure}

Figure~\ref{fig_dehmdp_throughput_distribution} and~\ref{fig_dehmdp_pktloss_distribution} depict the network throughput and packet loss rate of E-QAT with respect to $p_{i} (i \in [1,N])$ in Equation~(\ref{eq_probTx}). In Figure~\ref{fig_dehmdp_throughput_distribution}, $p_{i}$ following Exponential distribution with $\lambda = 0.5$ (shown as exp, $\lambda = 0.5$) achieves the highest network throughput, about 3000 packets when $N=40$. Moreover, the Gamma distribution of $p_{i}$ gives the lowest network throughput, which is about 2000 packets. In Figure~\ref{fig_dehmdp_pktloss_distribution}, when $N=10$, the four distributions perform similar packet loss rate, which is around 43\%. With an increase of number of nodes, Gamma distribution performs 7\% 30\%, and 40\% more packet loss than (exp, $\lambda = 1.5$), Sigmoid, and (exp, $\lambda = 0.5$), respectively. 

\begin{figure}[htb]
\centering
\includegraphics[width=4.5in]{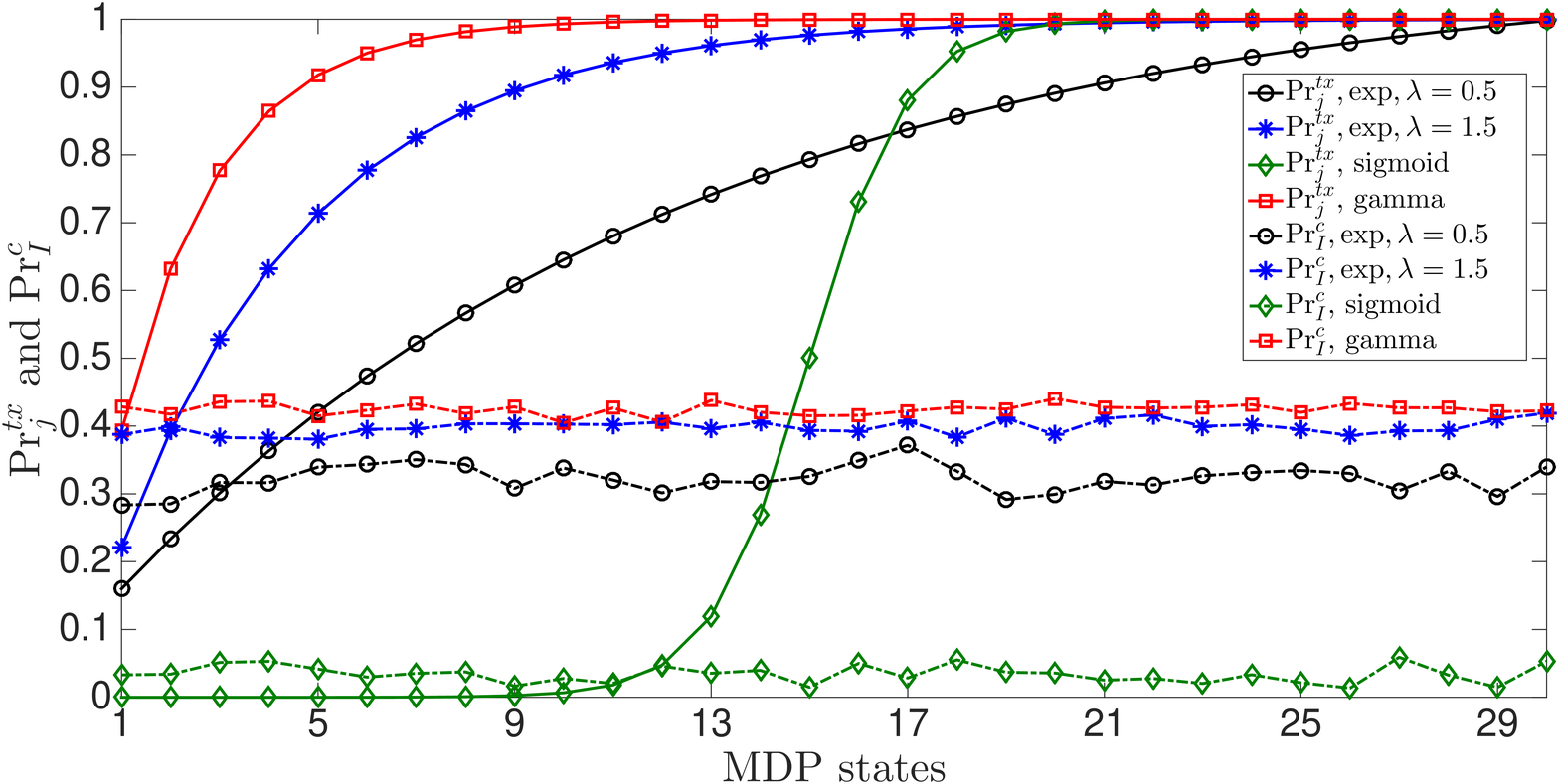}
\caption{\small{$p_{i} (i \in [1,N])$ and $\text{Pr}^{c}_{k}$ with different probability distributions.}}
\label{fig_probability_distribution}
\end{figure}

To further explain the reason of performance difference, Figure~\ref{fig_probability_distribution} plots the impact of different probability distributions to the $\text{Pr}^{tx}_{j}$ and $\text{Pr}^{c}_{k}$ over all MDP states. In total, the node has 30 MDP states since the maximum discretized energy level $E$ is 5 and queue length $Q$ is 6. Given the four probability distributions, $\text{Pr}^{tx}_{j}$ generally increases with the growth of the state, because the node with more energy or longer data queue has a higher transmission probability. 
We also observe in Figure~\ref{fig_probability_distribution} that $\text{Pr}^{c}_{k}$ stabilizes regarding to the states. Moreover, the Sigmoid distribution performs lower $\text{Pr}^{c}_{k}$ than the other three. The reason is that the $\text{Pr}^{tx}_{j}$ of Sigmoid distribution is lower than the others from state 1 to 16. Thus, when the nodes ($\neq k$) are in low MDP state, the $\text{Pr}^{c}_{k}$ has low value due to the definition of Equation~(\ref{eq_probCollision}). 

\subsubsection{Impact of $\widehat{T}$}
\begin{figure}[htb]
\centering
\includegraphics[width=4.5in]{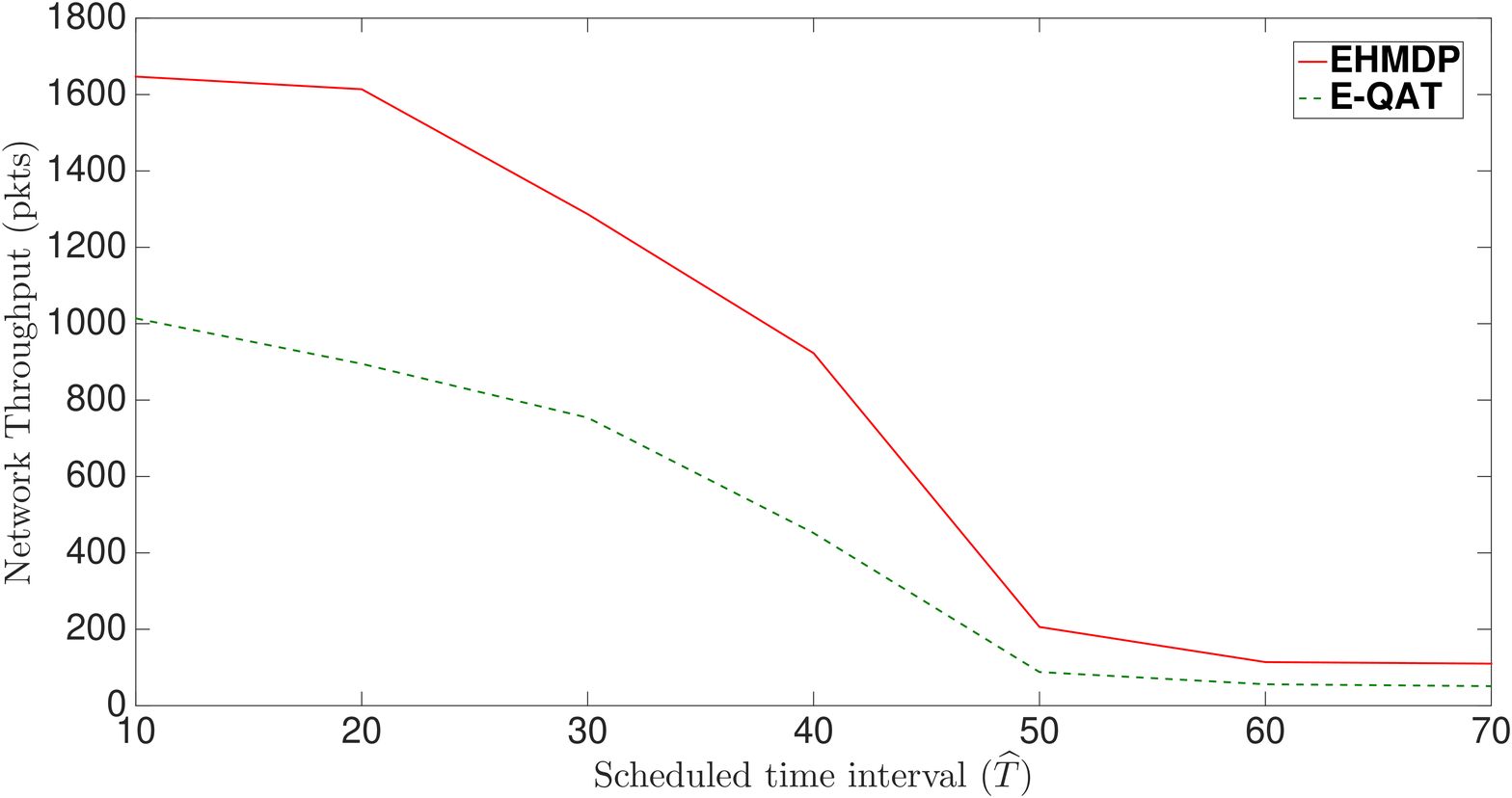}
\caption{Network throughput regarding to different $\widehat{T}$.}
\label{fig_throughput_interval}
\end{figure}

\begin{figure}[htb]
\centering
\includegraphics[width=4.5in]{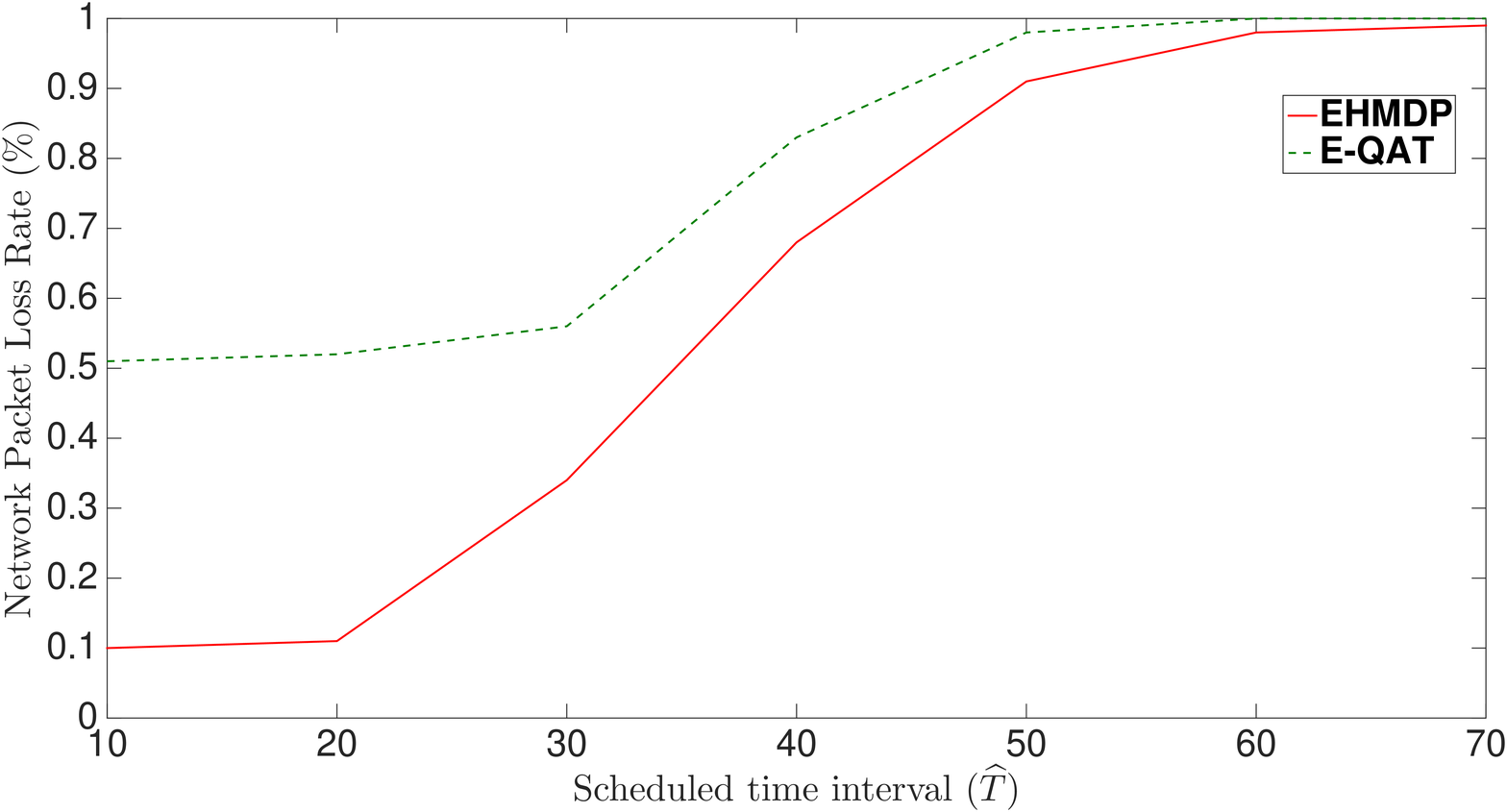}
\caption{Network packet loss regarding to different $\widehat{T}$.}
\label{fig_pktloss_interval}
\end{figure}

In this case, we study impact of the scheduling interval that lasts $\widehat{T}$ mini-slots. The number of nodes is fixed to 15, and the other settings are as configured in Section~\ref{simulation}-A. Figures~\ref{fig_throughput_interval} and~\ref{fig_pktloss_interval} present the network throughput and packet loss rate of EHMDP and E-QAT, where $\widehat{T} \in [10,40]$. 
It is observed that the network throughput of EHMDP and E-QAT generally drops with $\widehat{T}$ while the packet loss rates of EHMDP and E-QAT grow. It confirms the fact that an increasing number of packets is generated and injected into the data queues of nodes with a longer $\widehat{T}$. However, only one node is scheduled to transmit data and harvest energy during each $\widehat{T}$, though the node can harvest more energy with a longer $\widehat{T}$ due to~\eqref{eq_deltaE}. As a result, the nodes that are not scheduled suffer from more packet loss caused by data buffer overflow, compared to those with shorter $\widehat{T}$. Moreover, EHMDP achieves higher network throughput and lower packet loss than E-QAT, since the former has state information of each node and can accordingly achieve the optimal schedules. 

\section{Conclusion}
\label{cond}
In this paper, we address a scheduling problem on the data collection and WPT in RWSN. The problem is formulated as a finite state MDP with the objective of minimizing the packet loss. In terms of the scheduling mechanism of MAC protocol, two different MDP formulations have been studied, i.e., collision-free centralized scheduling, EHMDP, and contention-based semi-decentralized scheduling, E-QAT. Moreover, an off-line optimization of actions in centralized scheduling is investigated to provide the lower bound of packet loss to the scheduling algorithms of the WPT powered sensor networks. For the semi-decentralized scheduling, we develop a solution to obtain the optimal scheduling policy with different packet collision probability distributions.
Numerical results have shown that our proposed MDP formulation and algorithm outperform existing algorithms with substantial gains on throughput and packet loss.

\appendix
\section{[Optimization of $\rho_i$]}
Given the goal of the MDP to minimize the packet loss stemming from insufficient energy, $\rho_{i}$ is to be chosen to maximize the energy gained during each scheduled time interval, or ``epoch", with a duration of $\widehat{T}$. 
The optimal modulation of node $i$, $\rho_{i}$, is independent of the battery level and the queue length of the node $i$. This is because $\rho_i$ is selected to maximize the increase of the battery level at node $i$, not the battery level itself, under the bit error rate requirement $\epsilon_i$ for the packet transmitted. 
As a result, $\rho_{i}$ is decoupled from $\mathcal{A}$, and optimized a priori by
\begin{equation}
\begin{aligned}
\rho_i=&\arg\max_{\rho=1,\cdots,M} \bigg\{(\widehat{T}-\frac{L}{\rho W}) P_{i}^E -\frac{L}{\rho W}P_{i}^D(\rho)\bigg\}, \\
\end{aligned}
\label{eq_rho}
\end{equation}
the right-hand side (RHS) of which, by substituting \eqref{eq: transfer power} and \eqref{eq: transmit power}, can be rewritten as
\begin{equation} \label{eq: maxRho}
\begin{aligned}
&\max_{\rho=1,\cdots,M} \bigg\{(\widehat{T}-\frac{L}{\rho W}) P_{e} \|\mathbf{h}_{i}\|^2-\frac{L\kappa_2^{-1}\ln(\frac{\kappa_1}{\epsilon})}{\|\mathbf{h}_{i}\|^2\rho W}\big(2^{\rho}-1\big)\bigg\}
\end{aligned}
\end{equation}
where $W$ is the bandwidth of the uplink data transmission, $\frac{1}{W}$ is the duration of an uplink symbol, $\frac{L}{\rho W}$ is the duration of uplink data transmission, and $(\widehat{T}-\frac{L}{\rho W})$ is the rest of the epoch used for downlink WPT.

By using the first-order necessary condition of the optimal solution, we have 
\begin{equation}
\frac{d}{d\rho}((\widehat{T}-\frac{L}{\rho W}) P_{e} \|\mathbf{h}_{i}\|^2-\frac{L\kappa_2^{-1}\ln(\frac{\kappa_1}{\epsilon})}{\|\mathbf{h}_{i}\|^2\rho W}\big(2^{\rho}-1\big)) = 0
\end{equation}

\begin{equation}
\begin{aligned}
\rho^{-2}\frac{L}{W}P_{e} \|\mathbf{h}_{i}\|^2 &- \frac{L\kappa_2^{-1}\ln(\frac{\kappa_1}{\epsilon})}{\|\mathbf{h}_{i}\|^2 W}(\rho^{-1}2^{\rho}\text{In}2-\rho^{-2}2^{\rho}) \\
&- \frac{L\kappa_2^{-1}\ln(\frac{\kappa_1}{\epsilon})}{\|\mathbf{h}_{i}\|^2 W}\rho^{-2} = 0
\end{aligned}
\end{equation}

The $\rho$ values are then given as follows: 
\begin{equation}
\rho2^{\rho}\text{In}2-2^{\rho} = \frac{L}{W}P_{e} \|\mathbf{h}_{i}\|^2 \frac{\|\mathbf{h}_{i}\|^2 W}{L\kappa_2^{-1}\ln(\frac{\kappa_1}{\epsilon})} - 1 
\label{eq_rho_final}
\end{equation} 
Since the left-hand side of Equation~(\ref{eq_rho_final}) monotonically increases with $\rho$, the optimal value $\rho^{\star}$ is able to be obtained by applying a bisectional method, and evaluating the two closest integers about the fixed point of the bisectional method.

\bibliographystyle{elsarticle-num}
\bibliography{EHMDP}

\begin{thebibliography}{10}
\expandafter\ifx\csname url\endcsname\relax
  \def\url#1{\texttt{#1}}\fi
\expandafter\ifx\csname urlprefix\endcsname\relax\def\urlprefix{URL }\fi
\expandafter\ifx\csname href\endcsname\relax
  \def\href#1#2{#2} \def\path#1{#1}\fi

\bibitem{fu2016esync}
L.~Fu, L.~He, P.~Cheng, Y.~Gu, J.~Pan, J.~Chen, {ES}ync: Energy synchronized
  mobile charging in rechargeable wireless sensor networks, IEEE Transactions
  on Vehicular Technology 65~(9) (2016) 7415--7431.

\bibitem{che2015spatial}
Y.~L. Che, L.~Duan, R.~Zhang, Spatial throughput maximization of wireless
  powered communication networks, IEEE Journal on Selected Areas in
  Communications 33~(8) (2015) 1534--1548.

\bibitem{xiao2014wireless}
L.~Xiao, P.~Wang, D.~Niyato, D.~Kim, Z.~Han, Wireless networks with {RF} energy
  harvesting: A contemporary survey, IEEE Communications Surveys \& Tutorials
  17~(2) (2014) 757--789.

\bibitem{sudevalayam2011energy}
S.~Sudevalayam, P.~Kulkarni, Energy harvesting sensor nodes: Survey and
  implications, IEEE Communications Surveys \& Tutorials 13~(3) (2011)
  443--461.

\bibitem{zhang2016optimal}
Y.~Zhang, D.~Niyato, P.~Wang, D.~I. Kim, Optimal energy management policy of
  mobile energy gateway, IEEE Transactions on Vehicular Technology 65~(5)
  (2016) 3685--3699.

\bibitem{li2016fair}
K.~Li, C.~Yuen, B.~Kusy, R.~Jurdak, A.~Ignjatovic, S.~S. Kanhere, S.~Jha, Fair
  scheduling for data collection in mobile sensor networks with energy
  harvesting, arXiv preprint arXiv:1603.02476.

\bibitem{shu2016near}
Y.~Shu, H.~Yousefi, P.~Cheng, J.~Chen, Y.~J. Gu, T.~He, K.~G. Shin,
  Near-optimal velocity control for mobile charging in wireless rechargeable
  sensor networks, IEEE Transactions on Mobile Computing 15~(7) (2016)
  1699--1713.

\bibitem{li2015poster}
K.~Li, C.~Yuen, S.~Jha, Fair scheduling for energy harvesting {WSN} in smart
  city, in: Sensys, ACM, 2015, pp. 419--420.

\bibitem{wang2016adaptively}
Z.~Wang, L.~Duan, R.~Zhang, Adaptively directional wireless power transfer for
  large-scale sensor networks, IEEE Journal on Selected Areas in Communications
  34~(5) (2016) 1785--1800.

\bibitem{gong2016optimal}
S.~Gong, L.~Duan, N.~Gautam, Optimal scheduling and beamforming in relay
  networks with energy harvesting constraints, IEEE Transactions on Wireless
  Communications 15~(2) (2016) 1226--1238.

\bibitem{che2015multiantenna}
Y.~Che, J.~Xu, L.~Duan, R.~Zhang, Multiantenna wireless powered communication
  with cochannel energy and information transfer, IEEE Communications Letters
  19~(12) (2015) 2266--2269.

\bibitem{ding2015application}
Z.~Ding, C.~Zhong, D.~W.~K. Ng, M.~Peng, H.~A. Suraweera, R.~Schober, H.~V.
  Poor, Application of smart antenna technologies in simultaneous wireless
  information and power transfer, IEEE Communications Magazine 53~(4) (2015)
  86--93.

\bibitem{nan2016energy}
Z.~Nan, T.~Chen, X.~Wang, W.~Ni, Energy-efficient transmission schedule for
  delay-limited bursty data arrivals under nonideal circuit power consumption,
  IEEE Transactions on Vehicular Technology 65~(8) (2016) 6588--6600.

\bibitem{chen2015provisioning}
X.~Chen, W.~Ni, X.~Wang, Y.~Sun, Provisioning quality-of-service to energy
  harvesting wireless communications, IEEE Communications Magazine 53~(4)
  (2015) 102--109.

\bibitem{chen2016optimal}
X.~Chen, W.~Ni, X.~Wang, Y.~Sun, Optimal quality-of-service scheduling for
  energy-harvesting powered wireless communications, IEEE Transactions on
  Wireless Communications 15~(5) (2016) 3269--3280.

\bibitem{cui2017energy}
Q.~Cui, T.~Yuan, W.~Ni, Energy-efficient two-way relaying under non-ideal power
  amplifiers, IEEE Transactions on Vehicular Technology 66~(2) (2017)
  1257--1270.

\bibitem{cui2017energy2}
Q.~Cui, Y.~Zhang, W.~Ni, M.~Valkama, R.~J{\"a}ntti, Energy efficiency
  maximization of full-duplex two-way relay with non-ideal power amplifiers and
  non-negligible circuit power, IEEE Transactions on Wireless Communications
  16~(9) (2017) 6264--6278.

\bibitem{ni2013new}
W.~Ni, I.~B. Collings, A new adaptive small-cell architecture, IEEE Journal on
  Selected Areas in Communications 31~(5) (2013) 829--839.

\bibitem{he2013energy}
S.~He, J.~Chen, F.~Jiang, D.~K. Yau, G.~Xing, Y.~Sun, Energy provisioning in
  wireless rechargeable sensor networks, IEEE Transactions on Mobile Computing
  12~(10) (2013) 1931--1942.

\bibitem{fu2016optimal}
L.~Fu, P.~Cheng, Y.~Gu, J.~Chen, T.~He, Optimal charging in wireless
  rechargeable sensor networks, IEEE Transactions on Vehicular Technology
  65~(1) (2016) 278--291.

\bibitem{zhang2016data}
Y.~Zhang, S.~He, J.~Chen, Data gathering optimization by dynamic sensing and
  routing in rechargeable sensor networks, IEEE/ACM Transactions on Networking
  24~(3) (2016) 1632--1646.

\bibitem{zhao2016optimal}
F.~Zhao, L.~Wei, H.~Chen, Optimal time allocation for wireless information and
  power transfer in wireless powered communication systems, IEEE Transactions
  on Vehicular Technology 65~(3) (2016) 1830--1835.

\bibitem{xu2014multiuser}
J.~Xu, L.~Liu, R.~Zhang, Multiuser {MISO} beamforming for simultaneous wireless
  information and power transfer, IEEE Transactions on Signal Processing
  62~(18) (2014) 4798--4810.

\bibitem{zhou2014wireless}
X.~Zhou, R.~Zhang, C.~K. Ho, Wireless information and power transfer in
  multiuser ofdm systems, IEEE Transactions on Wireless Communications 13~(4)
  (2014) 2282--2294.

\bibitem{lee2016energy}
K.~Lee, J.-P. Hong, Energy-efficient resource allocation for simultaneous
  information and energy transfer with imperfect channel estimation, IEEE
  Transactions on Vehicular Technology 65~(4) (2016) 2775--2780.

\bibitem{kang2015full}
X.~Kang, C.~K. Ho, S.~Sun, Full-duplex wireless-powered communication network
  with energy causality, IEEE Transactions on Wireless Communications 14~(10)
  (2015) 5539--5551.

\bibitem{tekbiyik2013proportional}
N.~Tekbiyik, T.~Girici, E.~Uysal-Biyikoglu, K.~Leblebicioglu, Proportional fair
  resource allocation on an energy harvesting downlink, IEEE Transactions on
  Wireless Communications 12~(4) (2013) 1699--1711.

\bibitem{maravsevic2017max}
J.~Mara{\v{s}}evi{\'c}, C.~Stein, G.~Zussman, Max-min fair rate allocation and
  routing in energy harvesting networks: Algorithmic analysis, Algorithmica
  Journal 78~(2) (2017) 521--557.

\bibitem{qin2017joint}
C.~Qin, W.~Ni, H.~Tian, R.~P. Liu, Y.~J. Guo, Joint beamforming and user
  selection in multiuser collaborative {MIMO} {SWIPT} systems with
  non-negligible circuit energy consumption, IEEE Transactions on Vehicular
  Technology.

\bibitem{kunikawa2015fair}
M.~Kunikawa, H.~Yomo, K.~Abe, T.~Ito, A fair polling scheme for energy
  harvesting wireless sensor networks, in: Vehicular Technology Conference (VTC
  Spring), IEEE, 2015, pp. 1--5.

\bibitem{naderi2014rf}
M.~Y. Naderi, P.~Nintanavongsa, K.~R. Chowdhury, {RF-MAC}: A medium access
  control protocol for re-chargeable sensor networks powered by wireless energy
  harvesting, IEEE Transactions on Wireless Communications 13~(7) (2014)
  3926--3937.

\bibitem{nintanavongsa2013medium}
P.~Nintanavongsa, M.~Y. Naderi, K.~R. Chowdhury, Medium access control protocol
  design for sensors powered by wireless energy transfer, in: INFOCOM, IEEE,
  2013, pp. 150--154.

\bibitem{chen2013energy}
X.~Chen, X.~Wang, X.~Chen, Energy-efficient optimization for wireless
  information and power transfer in large-scale mimo systems employing energy
  beamforming, IEEE Wireless Communications Letters 2~(6) (2013) 667--670.

\bibitem{li2016energy}
K.~Li, W.~Ni, X.~Wang, R.~P. Liu, S.~S. Kanhere, S.~Jha, Energy-efficient
  cooperative relaying for unmanned aerial vehicles, IEEE Transactions on
  Mobile Computing 15~(6) (2016) 1377--1386.

\bibitem{he2014optimal}
T.~He, X.~Wang, W.~Ni, Optimal chunk-based resource allocation for {OFDMA}
  systems with multiple {BER} requirements, IEEE Transactions on Vehicular
  Technology 63~(9) (2014) 4292--4301.

\bibitem{kadrolkar2012variable}
A.~Kadrolkar, R.~X. Gao, R.~Yan, W.~Gong, Variable-word-length coding for
  energy-aware signal transmission, IEEE Transactions on Instrumentation and
  Measurement 61~(4) (2012) 850--864.

\bibitem{puterman2014markov}
M.~L. Puterman, Markov decision processes: discrete stochastic dynamic
  programming, John Wiley \& Sons, 2014.

\bibitem{arruda2010toward}
E.~F. Arruda, F.~Ourique, A.~Almudevar, Toward an optimized value iteration
  algorithm for average cost markov decision processes, in: IEEE Conference on
  Decision and Control (CDC), 2010, pp. 930--934.

\bibitem{sun2008constrained}
C.~Sun, E.~Stevens-Navarro, V.~W. Wong, A constrained mdp-based vertical
  handoff decision algorithm for 4g wireless networks, in: IEEE International
  Conference on Communications (ICC), IEEE, 2008, pp. 2169--2174.

\bibitem{chaturvedi2016design}
T.~Chaturvedi, K.~Li, C.~Yuen, A.~Sharma, L.~Dai, M.~Zhang, On the design of
  {MAC} protocol and transmission scheduling for internet of things, in: IEEE
  Region 10 Conference (TENCON), IEEE, 2016, pp. 2000--2003.

\bibitem{li2014kappa}
K.~Li, B.~Kusy, R.~Jurdak, A.~Ignjatovic, S.~S. Kanhere, S.~Jha,
  $\kappa$-{FSOM}: Fair link scheduling optimization for energy-aware data
  collection in mobile sensor networks, in: European Conference on Wireless
  Sensor Networks (EWSN), Springer, 2014, pp. 17--33.

\bibitem{ni2012q}
J.~Ni, B.~Tan, R.~Srikant, {Q-CSMA}: Queue-length-based {CSMA/CA} algorithms
  for achieving maximum throughput and low delay in wireless networks, IEEE/ACM
  Transactions on Networking 20~(3) (2012) 825--836.

\bibitem{shrestha2011markov}
B.~Shrestha, E.~Hossain, K.~W. Choi, S.~Camorlinga, A markov decision process
  ({MDP})-based congestion-aware medium access strategy for ieee 802.15. 4, in:
  Global Telecommunications Conference (GLOBECOM),, IEEE, 2011, pp. 1--5.

\bibitem{tay2004collision}
Y.~Tay, K.~Jamieson, H.~Balakrishnan, Collision-minimizing {CSMA} and its
  applications to wireless sensor networks, IEEE Journal on Selected Areas in
  Communications 22~(6) (2004) 1048--1057.

\end{thebibliography}

\end{document}